\documentstyle[12pt,epsfig]{article}

\begin{document}
\begin {center}
{\bf {\Large
$\pi^0$ and $\eta$ meson exchange interactions in the coherent $\eta$ meson
production in the nucleus} }
\end {center}
\begin {center}
Swapan Das \footnote {email: swapand@barc.gov.in} \\
{\it Nuclear Physics Division,
Bhabha Atomic Research Centre  \\
Mumbai-400085, India }
\end {center}

\begin {abstract}
The energy transfer spectra have been calculated for the coherent $\eta$
meson production in the proton nucleus reaction. The elementary reaction
occurring in the nucleus is considered as $ pN \to pN^*(1535) $;
$ N^*(1535) \to N\eta $. The $N^*(1535)$ excitation for the forward
going proton and $\eta$ meson is occurred due to the $\pi^0$ and $\eta$
meson exchange interactions, other mesons do not contribute in this
process.
The cross sections are calculated for the $\pi^0$ and $\eta$ meson exchange
interactions, and the effect of their interference on the cross section is
also studied. The initial and final state interactions as well as the
$N^*(1535)$ nucleus interaction are addressed by the optical potentials. The
distorted wave functions for proton and $\eta$ meson are described by the
eikonal form. The sensitivity of the cross section to the hadron nucleus
potential is presented.
\end {abstract}


PACS number(s): 25.40e, 13.30.Eg, 13.60.Le

\section{Introduction}

One of the current interests in the intermediate energy nuclear physics is
to explore the physics of $\eta$ meson \cite{wgpr}. Several data sets
for the $\eta$ meson production in the hadron induced reactions
have been available from various laboratories, like COSY \cite{cosy}
(see the references there in), SATURNE \cite{satu}, Los Alamos \cite{loal},
Brookhaven \cite{broo}. The $\eta$ meson production in the heavy-ion
collisions is reported from GSI \cite{gsi}. Due to the advent of the
high duty electron accelerators at Jefferson Laboratory, Bates, MAMI,
ELSA, ..... etc, good quality data for the photo- and electro- production
of the $\eta$ meson are also available \cite{peeta}. These accelerator
facilities, along with the newly developed sophisticated detecting systems,
provide ample scope to explore the dynamics of the $\eta$ meson in the
nuclear and particle reactions.

The study of the $\eta$ meson production reaction opens various avenues
to learn many exciting physics. The $\eta N$ scattering length near
threshold is large and attractive which predicts the existence of a new
type of hadronic atom, i.e., bound or quasi-bound eta mesic nucleus
\cite{etmn, oset1, cosy2}.
Large charge symmetry violation in the $\pi^0-\eta$ mixing \cite{pemx}
provides a way to estimate the mass difference between $u$ and $d$
quarks.
Being an isoscalar particle, the $\eta$ meson can excite a nucleon to a
$I=\frac{1}{2}$ nucleonic resonance. Specifically, the $\eta$ meson
strongly couples only to $N^*(1535)$, a
$ I(J^P) = \frac{1}{2} (\frac{1}{2}^+) $ nucleonic resonance.
Therefore, the $\eta$ meson production in the nuclear reaction is a
potential tool to investigate the propagation of the $N^*(1535)$
resonance in the nucleus, in addition to the study of the $\eta$ meson
nucleus interaction in the final state \cite{mosl, shrf}.

The $\eta$ meson can be produced through the hadronic interaction by
scattering of the pion or proton off the proton or nuclear target.
Theoretical studies on these reaction, as done by various authors
\cite{petp, lawl, vett}, show that the $\eta$ meson in the final state
arises due to the decay of $N^*(1535)$ resonance produced in the
intermediate state. Sometime back, Alvaredo and Oset \cite{oset} studied
the coherent $\eta$ meson production in the $(p,p^\prime)$ reaction in
the spin-isospin saturated nucleus:
$ p + A(gs) \to p^\prime + A(gs) + \eta $. The elementary reaction
in the nucleus proceeds as $ pN \to pN^* $; $ N^*(1535) \to N\eta $.
The $N^*(1535)$ resonance produced in the intermediate state due to the
$\eta$ meson (a psuedoscalar-isoscalar meson) exchange interaction only,
specifically, for the forward going proton and $\eta$ meson. The
contributions from other meson exchange interactions vanish in this
reaction \cite{oset}.

It could be mentioned that both $\pi$ and $\eta$ mesons are pseudoscalar
particles but pion can't contribute to above reaction since it is an
isovector meson and this reaction involves isoscalar nucleus.
Contrast to this, both $\pi$ and $\eta$ meson exchange interactions
contribute to $ p \to N^* (1535) $ excitation in the spin-saturated
isospin-one nucleus.
We, therefore, consider the coherent $\eta$ meson production in the
scalar-isovector nucleus (e.g., $^{14}$C) through the $(p,p^\prime)$
reaction, and study the contributions due to the $\pi^0$, $\eta$ meson
exchange interactions and the interference of these interactions.
The
diagrammatic presentation of the elementary reaction occurring in the
nucleus is presented in Fig.~1. It should be mentioned that the $\eta$
meson takes away almost the whole energy transferred to the nucleus, i.e.,
$ E_\eta \approx q_0 [ = E_p-E_{p^\prime} ] $, whereas the momentum of this
meson is adjusted by the recoiling nucleus.
The distorted wave functions of protons and $\eta$ meson are described
by the eikonal form. The optical potentials (appearing in the $N^*$
propagator and the distorted wave functions for protons) are evaluated using
the $``t\varrho({\bf r})"$ approximation. We take the $\eta$ meson nucleus
optical potential from Ref.~\cite{oset}.

\section{Formalism}

The Lagrangian densities describing the $\pi NN$, $\pi NN^*$, $\eta NN$
and $\eta NN^*$ interactions \cite{erwe0} are given by
\begin{eqnarray}
{\cal L}_{\pi NN~} &=&
-ig_\pi F_\pi(q^2) {\bar N} \gamma_5 {\bf \tau} N \cdot {\bf \pi} \nonumber \\
{\cal L}_{\pi NN^*} &=&
-ig^*_\pi F^*_\pi(q^2) {\bar N^*} {\bf \tau} N \cdot {\bf \pi} \nonumber \\
{\cal L}_{\eta NN~} &=&
-ig_\eta F_\eta(q^2) {\bar N} \gamma_5 N \eta \nonumber \\
{\cal L}_{\eta NN^*} &=&
-ig^*_\eta F^*_\eta(q^2) {\bar N^*} N \eta,
\label{lag}
\end{eqnarray}
where $g_{\pi(\eta)}$ and $g^*_{\pi(\eta)}$ denote the $\pi(\eta) NN$
and $\pi(\eta) NN^*$ coupling constants respectively. The values for them
are $g_\pi=13.4$, $g^*_\pi=0.71$, $g_\eta=7.93$ and $g^*_\eta=1.86$.
$F_{\pi(\eta)}$
and $F^*_{\pi(\eta)}$ are the $\pi(\eta) NN$ and $\pi(\eta) NN^*$ form
factors which are given by
\begin{equation}
F_M(q^2)=F^*_M(q^2) = \frac{\Lambda^2_M - m^2_M}{\Lambda^2_M - q^2};
~~~~~ (M=\pi^0,\eta).
\label{ffc}
\end{equation}
$ q^2 [ = q^2_0 - {\bf q}^2 ] $ is the four-momentum transfer from
$pp^\prime$ vertex to the nucleus, i.e., $ q_0 = E_p - E_{p^\prime} $
and $ {\bf q} = {\bf k}_p - {\bf k}_{p^\prime}  $.  The form factors are
normalized to unity when the mesons are in on-shell. Values for the length
parameters are $ \Lambda_\pi = 1.3 $ GeV and $ \Lambda_\eta = 1.5 $ GeV
\cite{oset1, mahe}. $m_M$ denotes the mass of the pseudoscalar meson:
$ m_{\pi^0} \simeq 135 $ MeV and $ m_\eta \simeq 547 $ MeV.

The $T$-matrix for the coherent $\eta$ production in the $(p,p^\prime)$
reaction on a nucleus can be written as
\begin{equation}
T_{fi} = \Gamma_{N^* \to N\eta} \Lambda_{N^*}
\sum_{M=\pi^0,\eta} {\tilde V}_M (q)
\int d{\bf r} \chi^{(-)*} ({\bf k}_\eta, {\bf r}) G_{N^*} (m, {\bf r})
\varrho_I ({\bf r}) \chi^{(-)*} ({\bf k}_{p^\prime}, {\bf r})
\chi^{(+)} ({\bf k}_p, {\bf r}),
\label{tmx1}
\end{equation}
with $ \Lambda_{N^*} = {\not k} + m_{N^*} $.
$\varrho_I ({\bf r})$ is the isospin averaged density distribution of
the nucleus.
$ \Gamma_{N^* \to N\eta}$ denotes the decay vertex:
$N^*(1535) \to N\eta$. ${\tilde V}_M (q)$ is the pseudoscalar meson
(i.e., $M \equiv \pi^0 ~\mbox{or}~ \eta$) exchange interaction between the
beam proton and the nucleon in the nucleus (see Fig.~1). It is given by
\begin{equation}
{\tilde V}_M (q) = \Gamma_{M NN^*} {\tilde G}_M (q^2)
                   \Gamma_{M pp^\prime}.
\label{psmp}
\end{equation}
$\Gamma_{M NN^*}$ and $\Gamma_{M pp^\prime}$ appearing in
this equation represent the $\pi(\eta) NN^*$ and $\pi(\eta) pp^\prime$
vertex factors. They are described by the Lagrangians expressed in
Eq.~(\ref {lag}). ${\tilde G}_M (q^2)$ in the above equation denotes the
propagator arising due to the virtual $\pi(\eta)$ meson exchange
interaction. The form for it is
\begin{equation}
{\tilde G}_{\pi(\eta)} (q^2) = - \frac{1}{m^2_{\pi(\eta)} - q^2}.
\label{vmpr}
\end{equation}

$\chi$s in Eq.~(\ref {tmx1}) represent the distorted wave function for the
projectile $p$, ejectile $p^\prime$ and $\eta$ meson. Using Glauber model
\cite{glub}, we can write $\chi$ for the beam proton $p$ \cite{das} as
\begin{equation}
\chi^{(+)} ({\bf k}_p, {\bf r}) = e^{i {\bf k}_p. {\bf r} }
exp[ -\frac{i}{v_p} \int^z_{-\infty} dz^\prime V_{Op} ({\bf b}, z^\prime) ],
\label{dwptn}
\end{equation}
where $v_p$ is the velocity of the proton. $V_{Op} ({\bf b}, z^\prime)$
describes the optical potential for it. For outgoing particles, i.e.,
$p^\prime$ and $\eta$ meson, the form for their distorted wave functions
is given by
\begin{equation}
\chi^{(-)^*} ({\bf k}_X, {\bf r}) = e^{-i {\bf k}_X. {\bf r} }
exp[ -\frac{i}{v_X} \int^{+\infty}_z dz^\prime V_{OX} ({\bf b}, z^\prime) ],
~~~~~ (X = p^\prime, \eta).
\label{dwpet}
\end{equation}
The optical potentials appearing in the above equations describe the
initial and final state interactions.

The propagator of the $N^*$ baryon, i.e., $G_{N^*} (m, {\bf r})$ in
Eq.~(\ref {tmx1}), can be expressed as
\begin{equation}
G_{N^*} (m, {\bf r}) = \frac{1}
{m^2-m^2_{N^*} + im_{N^*}\Gamma_{N^*}(m) - 2E_{N^*}V_{ON^*}({\bf r})},
\label{nspr}
\end{equation}
with $m_{N^*}=1.535$ GeV. $E_{N^*}$ is the energy of $N^*$ resonance. $m$
represents the invariant mass of the $\eta$ meson and nucleon, arising
due to the decay of $N^*$. $V_{ON^*}({\bf r})$ is the optical potential
which describes the interaction taking place between the $N^*$ resonance
and nucleus.
$\Gamma_{N^*}(m)$ denotes the total width of $N^*$ for its mass equal to
$m$. It consists of the partial decay widths due to the
$N^* \to N\pi (\sim 48\%)$, $N^* \to N\eta (\sim 42\%)$ and
$N^* \to N\pi\pi (\sim 10\%)$:
\begin{equation}
\Gamma_{N^*}(m) \approx \Gamma_{N^*}(m_{N^*}) \left [
  0.48\frac{k_\pi(m)}{k_\pi(m_{N^*})}
+ 0.42\frac{k_\eta(m)}{k_\eta(m_{N^*})} + 0.1 \right ];
~~~ \Gamma_{N^*}(m_{N^*}) = 0.15 ~ \mbox{GeV}.
\label{twdh}
\end{equation}
In this equation, $k_{\pi(\eta)}(m)$ is the pseudoscalar meson momentum
arising due to the $N^*$ of mass $m$ decaying at rest. It can be evaluated
using the equation:
\begin{equation}
k_M(m)
= \frac{ [ ( m^2 - (m_N+m_M)^2 )( m^2 - (m_N-m_M)^2 )]^{1/2} }{ 2m };
~~~~~ (M = \pi^0, \eta).
\label{kmsn}
\end{equation}

The differential cross section for the reaction illustrated above is
given by
\begin{equation}
\frac{d\sigma}{dE_{p^\prime}d\Omega_{p^\prime}d\Omega_\eta}
=K_F <|T_{fi}|^2>,
\label{dcrss}
\end{equation}
where the annular brackets around $|T_{fi}|^2$ represent the average over
the spins in the initial state and the summation over the spins in the
final state. $K_F$ is the kinematical factor for the reaction:
\begin{equation}
K_F = \frac{\pi}{(2\pi)^6}
\frac{ m^2_p m_A k_{p^\prime} k^2_\eta }
{ k_p | k_\eta (E_i-E_{p^\prime}) - E_\eta {\bf q} . k_\eta | }.
\label{kfc}
\end{equation}
All symbols carry their usual meanings.

\section{Results and Discussions}

The optical potential $V_{OX} ({\bf r})$, appearing in the $N^*$
propagator $ G_{N^*} (m, {\bf r}) $ in Eq.~(\ref {nspr}) and the
distorted wave functions $\chi$s in Eqs.~(\ref {dwptn}) and (\ref {dwpet}),
is calculated using the $ ``t\varrho ({\bf r})" $ approximation
\cite{das1}, i.e.,
\begin{equation}
V_{OX} ({\bf r})
= -\frac{v_X}{2} [i+\alpha_{XN}] \sigma^{XN}_t \varrho ({\bf r}),
\label{opts}
\end{equation}
where the symbol $X$ stands for the proton or $N^*(1535)$ resonance.
$v_X$ is the velocity of the particle $X$. $\alpha_{XN}$ denotes the ratio
of the real to imaginary part of the scattering amplitude $f_{XN}$.
$\sigma^{XN}_t$ represents the corresponding total cross section. To
evaluate the proton nucleus optical potential, i.e., $ V_{Op} ({\bf r}) $
as well as $ V_{Op^\prime} ({\bf r}) $, we use the energy dependent
experimentally determined values for $\alpha_{pN}$ and $\sigma^{pN}_t$
\cite{nndt}.
The measured values for the $N^*$ nucleon scattering parameters
$\alpha_{N^*N}$ and $\sigma^{N^*N}_t$ are not available. To estimate them,
we take $ \alpha_{N^*N} \approx \alpha_{pN} $ and $ \sigma^{N^*N}_{el}
\approx \sigma^{pN}_{el} $ since the elastic scattering dynamics of the
$N^*$ resonance can be assumed not much different compare to that of a
proton.
For
the reactive part of $\sigma^{N^*N}_t$, the dynamics of $N^*$ can be
thought same as that of a nucleon at its kinetic energy enhanced by
$\Delta m$, i.e., $ \sigma_r^{N*N} (T_{N^*N}) \approx $
$ \sigma_r^{NN} (T_{N^*N}+ \Delta m) $. Here, $\Delta m$ is the mass
difference between the resonance and nucleon. $T_{N^*N}$ is the total
kinetic energy in the $N^*N$ center of mass system \cite{jkl}.

The $\eta$ meson nucleus optical potential $V_{O\eta}$ ({\bf r}) can be
estimated from the $\eta$ meson self-energy $\Pi_\eta$ ({\bf r}) in the
nucleus evaluated by Alvaredo and Oset \cite{oset}:
\begin{eqnarray}
\Pi_\eta ({\bf r}) = 2E_\eta V_{O\eta} ({\bf r}) = g^{*2}_\eta
\frac{ \varrho ({\bf r}) }
{ m - m_{N^*} + \frac{i}{2}\Gamma_{N^*}(m) - V_{ON^*} ({\bf r})
                                            + V_{ON} ({\bf r}) },
\label{opet}
\end{eqnarray}
where the nucleon potential energy is taken as
$ V_{ON} ({\bf r}) = -50 \varrho ({\bf r}) / \varrho (0) $ MeV \cite{oset}.

The factor $ \varrho ({\bf r}) $, appearing in Eqs.~(\ref{tmx1}),
(\ref {opts}) and (\ref {opet}), represents the spatial density distribution
of the nucleus. The form of $ \varrho ({\bf r}) $ for $^{14}$C nucleus
(a scalar-isovector nucleus), as extracted from the electron scattering
data \cite{andt}, is given by
\begin{eqnarray}
\varrho ({\bf r})
= \varrho_0 [1+w(r/c)^2] e^{-(r/c)^2};
~~~ w=1.38, ~c=1.73 ~\mbox{fm}.     
\label{vrrh}
\end{eqnarray}
This density distribution is normalized to the mass number of the nucleus.

The isospin averaged nuclear density distribution $ \varrho_I ({\bf r}) $
in Eq.~(\ref{tmx1}) is related to $ \varrho ({\bf r}) $ as
\begin{equation}
\varrho_I ({\bf r})
= \left [ \frac{Z}{A} C_I(p) + \frac{A-Z}{A} C_I(n) \right ]
  \varrho ({\bf r}),
\label{vrhI}
\end{equation}
where $C_I(p)$ and $C_I(n)$ are the isospin matrix elements for the
proton and neutron respectively. $C_I(p) = +1$ and $C_I(n) = -1$ are
the values for $\pi^0$ exchange potential where as both of them are
equal to +1 for $\eta$ meson exchange potential.

We represent in Fig.~2 the calculated differential cross section
$ \frac{ d\sigma }{ dE_{p^\prime} d\Omega_{p^\prime} d\Omega_\eta } $
for the energy $E_\eta$ distribution of $\eta$ meson, produced coherently
in the $(p,p^\prime)$ reaction on $^{14}$C nucleus at 2.5 GeV.
In fact, this figure shows the plane wave results where the resonance
N(1535) nucleus interaction $V_{ON^*}$ is not included.
This
contribution to the cross section due to $\pi^0$ meson exchange potential
$V_\pi(q)$ (short-dash curve) is much lesser than that due to $\eta$
meson exchange potential $V_\eta(q)$ (long-dash curve). The interference
of these potentials is distinctly visible in this figure, see dot-dash
curve.

Fig.~3 describes the dependence of the calculated cross section on the
hadron nucleus potentials at 2.5 GeV. The contributions from both
$V_\pi(q)$ and $V_\eta(q)$ are incorporated in these results. The
dot-dashed curve shows the plane wave ($V_{ON^*}$ not included) results
where as the solid curve elucidates the distorted wave ($V_{ON^*}$ included)
results. This figure shows the cross section is drastically reduced because
of the hadron nucleus potentials.

\section{Conclusions}

We have calculated the differential cross sections for the energy
distribution of $\eta$ meson which is produced coherently in the proton
nucleus (scalar-isovector) reaction. The $\eta$ meson in the final state
appears because of the decay of $N^*(1535)$ produced in the intermediate
state.
The
$N^*$ excitation in the nucleus is occurred due to the $\pi^0$ and $\eta$
meson exchange interactions between the projectile proton and a nucleon
inside the nucleus. The contribution of the pion exchange interaction to
the cross section is much smaller than that of the $\eta$ meson exchange
interaction.
The
cross section is drastically reduced because of the hadron nucleus
interactions.


\newpage

{\bf Figure Captions}
\begin{enumerate}

\item
(color online).
Schematic diagram of the elementary reaction:
$ pN \to p^\prime N^* ; ~N^* \to N\eta $.

\item
(color online).
The $\eta$ meson energy $E_\eta$ distribution spectra for $^{14}$C nucleus
at 2.5 GeV beam energy. The contribution due to the $\pi^0$ meson exchange
interaction $V_\pi(q)$ is shown by the short-dash curve whereas the
long-dash curve represents that due to the $\eta$ meson exchange
interaction $V_\eta(q)$. The cross section due to these interactions
(coherently added) is illustrated by the dot-dash curve.

\item
(color online).
The plane wave ($N^*$ potential $V_{ON^*}$ not included) results are
compared with the distorted wave ($V_{ON^*}$ included) results. The peak
cross section is reduced drastically because of the hadron nucleus
potentials.

\end{enumerate}

\newpage
\begin{figure}[h]
\begin{center}
\centerline {\vbox {
\psfig{figure=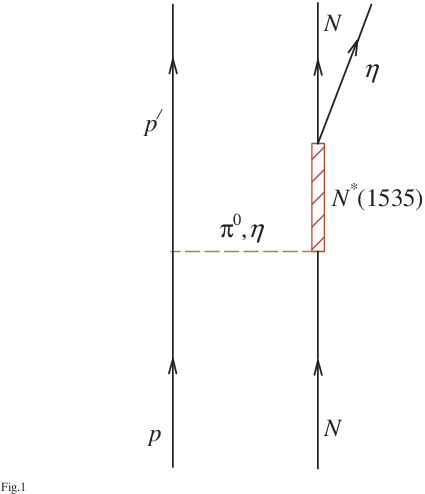,height=7.0 cm,width=6.0 cm}
}}
\caption{
(color online).
Schematic diagram of the elementary reaction:
$ pN \to p^\prime N^* ; ~N^* \to N\eta $.
}
\label{fig1}
\end{center}
\end{figure}

\newpage
\begin{figure}[h]
\begin{center}
\centerline {\vbox {
\psfig{figure=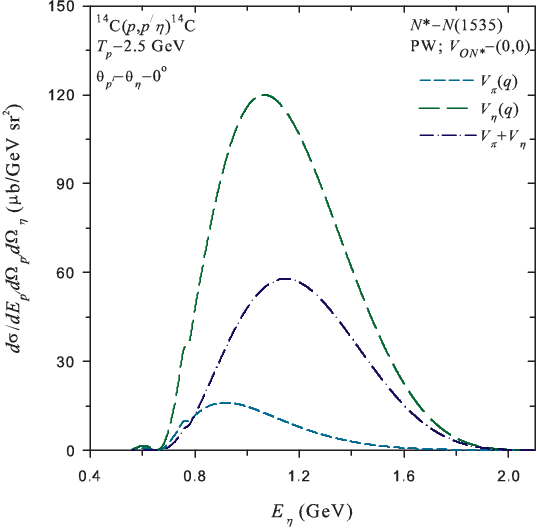,height=10.0 cm,width=10.0 cm}
}}
\caption{
(color online).
The $\eta$ meson energy $E_\eta$ distribution spectra for $^{14}$C nucleus
at 2.5 GeV beam energy. The contribution due to the $\pi^0$ meson exchange
interaction $V_\pi(q)$ is shown by the short-dash curve whereas the
long-dash curve represents that due to the $\eta$ meson exchange
interaction $V_\eta(q)$. The cross section due to these interactions
(coherently added) is illustrated by the dot-dash curve.
}
\label{fig2}
\end{center}
\end{figure}

\newpage
\begin{figure}[h]
\begin{center}
\centerline {\vbox {
\psfig{figure=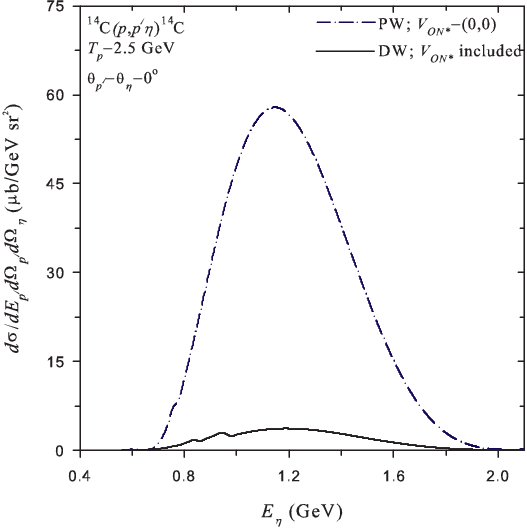,height=10.0 cm,width=10.0 cm}
}}
\caption{
(color online).
The plane wave ($N^*$ potential $V_{ON^*}$ not included) results are
compared with the distorted wave ($V_{ON^*}$ included) results. The peak
cross section is reduced drastically because of the hadron nucleus
potentials.
}
\label{fig3}
\end{center}
\end{figure}

\end{document}